\documentclass[seceq]{ptptex}

\usepackage{graphics}
\usepackage{epsfig}

\title{Multi-messenger observations of neutron rich matter}

\author{C. J. Horowitz}

\inst{Departement of Physics, University of Tennessee, and Physics Division, Oak Ridge National Laboratory, and \\ Center for the Exploration of Energy and Matter and Department of Physics, Indiana University, Bloomington, IN 47405, USA}
\abst{  
At very high densities, electrons react with protons to form neutron rich matter.  This material is central to many fundamental questions in nuclear physics and astrophysics.  Moreover, neutron rich matter is being studied with an extraordinary variety of new tools such as the Facility for Rare Isotope Beams  (FRIB) and the Laser Interferometer Gravitational Wave Observatory (LIGO).  We describe the Lead Radius Experiment (PREX) that uses parity violating electron scattering to measure the neutron radius in $^{208}$Pb.  This has important implications for neutron stars and their crusts. We discuss X-ray observations of neutron star radii.  These also have important implications for neutron rich matter.  Gravitational waves (GW) open a new window on neutron rich matter.  They come from sources such as neutron star mergers, rotating neutron star mountains, and collective r-mode oscillations.  Using large scale molecular dynamics simulations, we find neutron star crust to be very strong.  It can support mountains on rotating neutron stars large enough to generate detectable gravitational waves.  
We believe that combing astronomical observations using photons, GW, and neutrinos, with laboratory experiments on nuclei, heavy ion collisions, and radioactive beams will fundamentally advance our knowledge of compact objects in the heavens, the dense phases of QCD, the origin of the elements, and of neutron rich matter.  }

\begin{document}

\maketitle

\section{Introduction}

Multi-messenger astronomy observes matter under extreme conditions.  In this paper we describe how electromagnetic, gravitational wave, and neutrino astronomy, along with laboratory experiments, provide complimentary information on neutron rich matter.  Compress almost anything to very high densities and electrons react with protons to form neutron rich matter.  This material is at the heart of many fundamental questions in Nuclear Physics and Astrophysics.
\begin{itemize}
\item What are the high density phases of QCD?
\item Where did the chemical elements come from?
\item What is the structure of many compact and energetic
objects in the heavens, and what determines their
electromagnetic, neutrino, and gravitational-wave
radiations?
\end{itemize}
Furthermore, neutron rich matter is being studied with an extraordinary
variety of new tools such as the Facility for Rare Isotope
Beams (FRIB), a heavy ion accelerator to be built at
Michigan State University \cite{frib}, and the Laser Interferometer
Gravitational Wave Observatory (LIGO) \cite{ligo}.  Indeed there are many, qualitatively different, probes of neutron rich matter including precision laboratory measurements on stable nuclei and experiments with neutron rich radioactive beams.  While astrophysical observations probe neutron rich matter with electromagnetic radiation, neutrinos, and gravitational waves.   In this paper we give brief examples of how neutron rich matter is being studied with these extraordinarily different probes.

We are interested in neutron rich matter over a tremendous range of densities and temperatures were it can be a gas, a liquid, a solid, a plasma, a liquid crystal, a superconductor, a superfluid, a color superconductor, etc.  Neutron rich matter is a remarkably versatile material.  The liquid crystal phases are known as nuclear pasta and arise because of coulomb frustration \cite{pasta,watanabe}.  Pasta is expected at the base of the crust in a neutron star and can involve complex shapes such as long rods (``spaghetti'') or flat plates (``lasagna'').  Neutrinos in core collapse supernovae may scatter coherently from these shapes (neutrino pasta scattering) because the shapes have sizes comparable to the neutrino wavelength \cite{pastascattering}.  

In this paper we focus on some of the simpler gas, solid, and liquid phases of neutron rich matter.    In Section \ref{sec.nuclei} we describe a precision laboratory experiment called PREX to measure the neutron radius of $^{208}$Pb.  Nuclei are liquid drops, so PREX and many other laboratory experiments probe the {\it liquid} phase of neutron rich matter.   In astrophysics, electromagnetic, gravitational wave, and neutrino probes can observe different phases of neutron rich matter because the probes have very different mean free paths.  In Section \ref{sec.EM} we describe electromagnetic observations of neutron star radii.  In Section \ref{sec.GW} we discuss gravitational waves from neutron star mergers that are produced by the energetic motions of dense {\it liquid} phase neutron rich matter.   In addition, continuous gravitational waves can be produced by  ``mountains'' of {\it solid} neutron rich matter on rapidly rotating stars.  We conclude in Section \ref{sec.conclusions}.

\section{Laboratory probes of neutron rich matter}
\label{sec.nuclei}
Neutron rich matter can be studied in the laboratory.  Hot and or dense matter can be formed in heavy ion collisions, while more neutron rich conditions can be accessed with radioactive beams.  In addition precise experiments are possible on stable neutron rich nuclei.  We give one example, the Lead Radius Experiment (PREX) \cite{PREXI} accurately measures the neutron radius in $^{208}$Pb with parity violating electron scattering \cite{bigprex}.  This has many implications for nuclear structure, astrophysics, atomic parity violation, and low energy tests of the standard model. 

\subsection{Introduction to neutron densities and neutron radii}  

Nuclear charge densities have been accurately measured with electron scattering and have become our picture of the atomic nucleus, see for example ref. \cite{chargeden}.  These measurements have had an enormous impact.  
In contrast, our knowledge of neutron densities comes primarily from hadron scattering experiments involving for example pions \cite{pions}, protons \cite{protons1,protons2,protons3}, or antiprotons \cite{antiprotons1,antiprotons2}.  See also ref. \cite{tamii} for a beautiful measurement of the dipole polarizability.  However, the interpretation of hadron scattering experiments is model dependent because of uncertainties in the strong interactions.

Parity violating electron scattering provides a model independent probe of neutron densities that is free from most strong interaction uncertainties.  This is because the weak charge of a neutron is much larger than that of a proton \cite{dds}.  Therefore the $Z^0$ boson, that carries the weak force, couples primarily to neutrons.  In Born approximation, the parity violating asymmetry $A_{pv}$, the fractional difference in cross sections for positive and negative helicity electrons, is proportional to the weak form factor.  This is very close to the Fourier transform of the neutron density.  Therefore the neutron density can be extracted from an electro-weak measurement \cite{dds}.  
Many details of a practical parity violating experiment to measure neutron densities have been discussed in a long paper \cite{bigprex}.

The neutron radius of $^{208}$Pb, $R_n$, has important implications for astrophysics.  There is a strong correlation between $R_n$ and the pressure of neutron matter $P$ at densities near 0.1 fm$^{-3}$ (about 2/3 of nuclear density) \cite{alexbrown}.  A larger $P$ will push neutrons out against surface tension and increase $R_n$.  Therefore measuring $R_n$ constrains the equation of state (EOS) --- pressure as a function of density --- of neutron matter.  

Recently Hebeler et al. \cite{hebeler} used chiral perturbation theory to calculate the EOS of neutron matter including important contributions from very interesting three neutron forces.  
From their EOS, they predict $R_n-R_p= 0.17 \pm 0.03$ fm.  Here $R_p$ is the known proton radius of $^{208}$Pb.   Monte Carlo calculations by Carlson et al. also find sensitivity to three neutron forces \cite{MC3n}.   Therefore, measuring $R_n$ provides an important check of fundamental neutron matter calculations, and constrains three neutron forces.

The correlation between $R_n$ and the radius of a neutron star $r_{NS}$ is also very interesting \cite{rNSvsRn}.  In general, a larger $R_n$ implies a stiffer EOS, with a larger pressure, that will also suggest $r_{NS}$ is larger.  Note that this correlation is between objects that differ in size by 18 orders of magnitude from $R_n\approx 5.5$ fm to $r_{NS}\approx 10$ km.  We discuss observations of $r_{NS}$ in Section \ref{sec.EM}. 
 

The EOS of neutron matter is closely related to the symmetry energy $S$.  
This describes how the energy of nuclear matter rises as one goes away from equal numbers of neutrons and protons.  There is a strong correlation between $R_n$ and the density dependence of the symmetry energy $dS/dn$, with $n$ the baryon density.  The symmetry energy can be probed in heavy ion collisions \cite{isospindif}.  For example, $dS/dn$ has been extracted from isospin diffusion data \cite{isospindif2} using a transport model.

The symmetry energy $S$ helps determine the composition of a neutron star.     A large $S$, at high density, implies a large proton fraction $Y_p$ that will allow the direct URCA process of rapid neutrino cooling.  If $R_n-R_p$ is large, it is likely that massive neutron stars will cool quickly by direct URCA  \cite{URCA}.  In addition, the transition density from solid neutron star crust to the liquid interior is strongly correlated with $R_n-R_p$ \cite{cjhjp_prl}.  

Finally, atomic parity violation (APV) is sensitive to $R_n$ \cite{pollockAPV},\cite{brownAPV},\cite{bigprex}.  Parity violation involves the overlap of atomic  electrons with the weak charge of the nucleus, and this is primarily carried by the neutrons.  Furthermore, because of relativistic effects the electronic wave function can vary rapidly over the nucleus.  Therefore, the APV signal depends on where the neutrons are and on $R_n$.   A future low energy test of the standard model may involve the combination of a precise APV experiment along with PV electron scattering to constrain $R_n$.  Alternatively, measuring APV for a range of isotopes can provide information on neutron densities \cite{berkeleyAPV}.

\subsection{The Lead Radius Experiment (PREX)}

We now discuss a direct measurement of $R_n$.  Parity violation provides a model independent probe of neutrons, because the $Z^0$ boson couples to the weak charge, and the weak charge of a proton $Q_W^p=1-4\sin^2\Theta_W \approx 0.05$
is much smaller than the weak charge of a neutron $Q_W^n=-1$.
Here $\Theta_W$ is the weak mixing angle.  

The Lead Radius Experiment (PREX) at Jefferson Laboratory \cite{PREXI} measures the parity violating asymmetry $A_{pv}$ for elastic electron scattering from $^{208}$Pb.  The asymmetry $A_{pv}$ is the fractional cross section difference for scattering positive (+), or negative (-), helicity electrons,
\begin{equation}
A_{pv}=\frac{\frac{d\sigma}{d\Omega}|_+-\frac{d\sigma}{d\Omega}|_-}{\frac{d\sigma}{d\Omega}|_++\frac{d\sigma}{d\Omega}|_-}\, .
\end{equation}
In Born approximation, $A_{pv}$ arrises from the interference of a weak amplitude of order the Fermi constant $G_F$, and an electromagnetic amplitude of order the fine structure constant $\alpha$ over the square of the momentum transfer $q^2$ \cite{dds},
\begin{equation}
A_{pv}\approx \frac{G_F q^2\, F_W(q^2)}{2 \pi\alpha\sqrt{2} \, F_{ch}(q^2)}\, .
\label{apvborn}
\end{equation}
Here the weak form factor $F_W(q^2)$ is the Fourier  transform of the weak charge density $\rho_W(r)$, that is essentially the neutron density,
$F_W(q^2)=\int d^3r \frac{\sin(qr)}{qr} \rho_W(r)$.
Likewise, the electromagnetic form factor $F_{ch}(q^2)$ is the Fourier transform of the (electromagnetic) charge density $\rho_{ch}(r)$.  This is known from elastic electron scattering \cite{chargeden}. Therefore, measuring $A_{pv}$ as a function of $q$ allows one to map out the neutron density $\rho_n(r)$.  Note that, for a heavy nucleus, there are important corrections to Eq. \ref{apvborn} from Coulomb distortions.  However, these have been calculated exactly by solving the Dirac equation for an electron moving in both a Coulomb potential of order 25 MeV and a weak axial vector potential of order electron volts \cite{couldist}.  Therefore, even with Coulomb distortions, one can accurately determine neutron densities.   Note that this purely electroweak reaction is free from most strong interaction uncertainties.

The PREX experiment measures $A_{pv}$ for 1.05 GeV electrons elastically scattered from $^{208}$Pb at laboratory angles near five degrees.  
The first measurement yielded $A_{pv}=0.656\pm 0.060$ (statitistical) $\pm 0.014$ (systematic) ppm \cite{PREXI}.  From this the rms neutron radius $R_n$ minus proton radius $R_p$ for $^{208}$Pb was found to be $R_n-R_p=0.33^{+0.16}_{-0.18}$ fm.  See also ref. \cite{weakFF} for more details of this analyis.  A second PREX run is now approved to accumulate more statistics and reach the original goal of determining $R_n$ to 1\% ($\pm 0.05$ fm).

In addition to PREX, many other parity violating measurements of neutron densities are possible, see for example \cite{PREXII}.  Measuring $R_n$ in $^{48}$Ca is particularly attractive.  First, $^{48}$Ca has a higher experimental figure of merit than $^{208}$Pb.  Therefore a $^{48}$Ca measurement may take less beam time than for $^{208}$Pb.  Not only does $^{48}$Ca have a large neutron excess, it is also relatively light.  With only 48 nucleons, microscopic coupled cluster calculations \cite{coupledcluster}, or no core shell model calculations \cite{NCSM}, may be feasible for $^{48}$Ca that are presently not feasible for $^{208}$Pb.   Note that these microscopic calculations may have important contributions from three nucleon forces.  This will allow one to make microscopic predictions for the neutron density and relate a measured $R_n$ to three nucleon forces and in particular to very interesting three neutron forces.  


\section{Electromagnetic observations of neutron star radii}
\label{sec.EM}


The structure of a neutron star can be calculated with the Tolman-Oppenheimer-Volkoff Equations of General Relativity \cite{TOV} and is completely determined by the equation of state of neutron rich matter.  The radius of a neutron star depends on the pressure of neutron matter at normal nuclear density and above, because the central density of a neutron star can be a few or more times that of normal nuclear density.   A higher pressure will lead to a larger radius.  It is important to have both low density information on the equation of state from PREX, and high density information from measurements of neutron star radii.  This can constrain any possible density dependence of the equation of state from an interesting phase transition to a possible high density exotic phase such as quark matter, strange matter, or a color superconductor.  For example, if the $^{208}$Pb radius is relatively large, this shows the EOS is stiff at low density (has a high pressure).  If at the same time, neutron stars have relatively small radii, than the high density EOS is soft with a low pressure.  This softening of the EOS with density could strongly suggest a phase transition to a soft high density exotic phase.

The radius of a neutron star $r_{NS}$ can be deduced from X-ray measurements of luminosity $L$ and surface temperature $T$,
$L=4\pi r_{NS}^2 \sigma_{SB}T^4$,
with $\sigma_{SB}$ the Stefan Boltzmann constant.  
Recently Steiner, Lattimer, and Brown have deduced masses and radii \cite{SLB} from combined observations of six neutron stars in two classes: 1) X-ray bursts, and 2) neutron stars in globular clusters.  They conclude that observations favor a stiff high density equation of state that can support neutron stars with a maximum mass near 2 $M_\odot$ and that the equation of state is soft at low densities so that a 1.4 $M_\odot$ neutron star has a radius near 12 km.  They go on to predict that the neutron minus proton root mean square radius in $^{208}$Pb will be $R_n-R_p=0.15\pm 0.02$ fm.  Note that this is a prediction for a nucleus based on an equation of state deduced from X-ray observations of neutron stars.  The Steiner et al. paper \cite{SLB} is potentially controversial because their results depend on, among other things, the model assumed for X-ray bursts.  

\section{Gravitational Waves}
\label{sec.GW}

We turn now to gravitational wave observations of neutron rich matter.   Albert Einstein, almost 100 years ago, predicted the oscillation of space and time known as gravitational waves (GW).  Within a few years, with the operation of Advanced LIGO \cite{advancedLIGO}, Advanced VIRGO \cite{advancedVIRGO} and other sensitive interferometers, we anticipate the historic detection of GW.  This will be a remarkable achievement and open a new window on the universe and on neutron rich matter.  

The first GW that are detected will likely come from the merger of two neutron stars.  The rate of such mergers can be estimated from known binary systems \cite{LIGOrate}.   During a merger the GW signal has a so called chirp form where the frequency rises as the two neutron stars spiral closer together.  Deviations of this wave form from that expected for two point masses may allow one to deduce the equation of state of neutron rich matter and measure the radius of a neutron star $r_{NS}$ \cite{GWEOS}.   Alternatively one may be able to observe the frequency of oscillations of the hyper-massive neutron star just before it collapses to a black hole.  This frequency depends on the radius of the maximum mass neutron star \cite{Rmax}.  

Continuous GW signals can also be detected, see for example \cite{Collaboration:2009rfa}.
Indeed Bildstein and others \cite{bildstein} have speculated that some neutron stars in binary systems may radiate angular momentum in continuous GW at the same rate that angular momentum is gained from accretion.  This would explain why the fastest observed neutron stars are only spinning at about half of the breakup rate.  There are several very active ongoing and near future searches for continuous gravitational waves at LIGO, VIRGO and other detectors, see for example \cite{abbot}.  
No signal has yet been detected.  However, sensitive upper limits have been set.  These limits constrain the shape of neutron stars.  In some cases the star's elipticity $\epsilon$, which is that fractional difference in moments of inertia $\epsilon=(I_1-I_2)/I_3$ is observed to be less than a part per million or even smaller.  Here $I_1$, $I_2$, and $I_3$ are the principle moments of inertia.    

           

An asymmetric mass on a rapidly rotating neutron star produces a time dependent mass quadrupole moment that radiates gravitational waves.  However, one needs a way (strong stick) to hold the mass up.  Magnetic fields can support mountains, see for example \cite{magneticmountains}.  However, it may require large internal magnetic fields.  Furthermore, if a star also has a large external dipole field, electromagnetic radiation may rapidly spin the star down and reduce the GW radiation.  

Alternatively, mountains can be supported by the solid neutron star crust.
Recently we performed large scale MD simulations of the strength of neutron star crust \cite{crustbreaking,chugunov}.   A strong crust can support large deformations or ``mountains'' on neutron stars, see also \cite{lowmassNS}, that will radiate strong GW.   How large can a neutron star mountain be before it collapses under the extreme gravity?  This depends on the strength of the crust.  We performed large scale MD simulations of crust breaking, where a sample was strained by moving top and bottom layers of frozen ions in opposite directions \cite{crustbreaking}.  These simulations involve up to 12 million ions and explore the effects of defects, impurities, and grain boundaries on the breaking stress.  

We find that neutron star crust is very strong because the high pressure prevents the formation of voids or fractures and because the long range coulomb interactions insure many redundant ``bounds'' between planes of ions.  Neutron star crust is the strongest material known, according to our simulations.  The breaking stress is 10 billion times larger than that for steel.  This is very promising for GW searches because it shows that large mountains are possible, and these could produce detectable signals.

To conclude this section, there is a great deal of interest in gravitational waves (GW) from neutron stars and there are many ongoing searches.  One is interested in both burst sources, for example from neutron star mergers, and  continuous sources from mountains or collective modes.  Gravitational wave radiation depends on the equation of state of neutron rich matter.  In addition, it can also depend on other more detailed properties including the breaking strain of solid phases and the bulk and shear viscosities.

\section{Conclusions: neutron rich matter}
\label{sec.conclusions}
Neutron rich matter is at the heart of many fundamental questions in Nuclear
Physics and Astrophysics. What are the high density phases of QCD? Where did the chemical elements come from? What is the structure of many compact and energetic objects in the heavens, and what determines their electromagnetic, neutrino, and gravitational-wave radiations? Moreover, neutron rich matter is being studied with an extraordinary variety of new tools such as the Facility for Rare Isotope Beams (FRIB) and the Laser Interferometer Gravitational Wave Observatory (LIGO). 

We described the Lead Radius Experiment (PREX) that uses parity violating electron scattering to measure the neutron radius in $^{208}$Pb. This has important implications for neutron stars and their crusts. We discussed X-ray observations of neutron star radii that also have important implications for neutron rich matter.  Gravitational waves (GW) from sources such as neutron star mergers and rotating neutron star mountains open a new window on neutron rich matter.  Using large scale molecular dynamics simulations, we found neutron star crust to be the strongest material known, some 10 billion times stronger than steel.  It can support mountains on rotating neutron stars large enough to generate detectable gravitational waves.  

In conclusion, multi-messenger astronomy is based on the widely held belief that combining astronomical observations using photons, gravitational waves, and neutrinos will fundamentally advance our knowledge of compact and energetic objects in the heavens.  Compact objects such as neutron stars are, in fact, giant nuclei, even if they are an extraordinary 18 orders of magnitude larger than a $^{208}$Pb nucleus.  Nevertheless, both in the laboratory and in Astrophysics, these objects are made of the same neutrons, that undergo the same strong interactions, and have the same equation of state.  A measurement in one domain, be it Astrophysics or the laboratory,  can have important implications in the other domain.  Therefore we can generalize multi-messenger astronomy  to multi-messenger observations of neutron rich matter.   We believe that combing astronomical observations using photons, GW, and neutrinos, with laboratory experiments on nuclei, heavy ion collisions, and radioactive beams will fundamentally advance our knowledge of the heavens, the dense phases of QCD, the origin of the elements, and of neutron rich matter.

\section*{Acknowledgments}

This work was done in collaboration with many people including D. K. Berry, E. F. Brown, K. Kadau, J. Piekarewicz, and graduate students Liliana Caballero, Helber Dusan, Joe Hughto, Justin Mason, Andre Schneider and Gang Shen.  This work was supported in part by DOE grant DE-FG02-87ER40365 and by the National Science Foundation, TeraGrid grant TG-AST100014.

\end{document}